%% file: Main_short.tex
\documentclass[letterpaper, 10 pt, conference]{ieeeconf}

\input{packages}
\input{my_commands}

\usepackage{comment}
\IEEEoverridecommandlockouts

\title{\LARGE \bf
Towards Interaction Regulation from Human Feedback via Free Energy Minimization
}

\author{M. Paula Diaz Monfort\textsuperscript{1,*}, Cinzia Tomaselli\textsuperscript{1,*}, Michael Richardson\textsuperscript{2,\dag}, Giovanni Russo\textsuperscript{3,\dag} 
\thanks{\textsuperscript{1} Scuola Superiore Meridionale, Napoli, Italy, \textsuperscript{2} Macquarie University, Sydney, Australia, \textsuperscript{3} University of Salerno, Salerno, Italy. \textsuperscript{*} Joint first. \textsuperscript{\dag} Joint last. Corresponding: {\tt\small c.tomaselli@ssmeridionale.it 
    giovarusso@unisa.it }}%
}

\begin{document}
\maketitle
\thispagestyle{empty}
\pagestyle{empty}

\begin{abstract}
A central challenge across control and learning is the design of mechanisms regulating the interactions between humans and autonomous agents. Inspired by the free energy principle from computational neuroscience, we introduce a control-theoretical framework to integrate human preferences online into an agent policy. We turn the framework into an open control architecture and validate our approach using a human-in-the-loop experimental testbed involving a rover navigating via onboard sensing. The human, remotely located and equipped with virtual reality headsets, shares the same sensory information as the rover. Human preferences are provided to the rover via gestures which introduce both cooperative and competitive interactions between the agent goal and the preferences. The experiments show that interactions are regulated, validating the proposed approach. 
\end{abstract}

\section{Introduction}
A popular paradigm to design autonomous agents involves tuning the agent policy using information from humans. This paradigm is often associated with reinforcement learning from human feedback, where human preferences are typically used offline to learn reward models that are then optimized by the agent~\cite{NL:25}. In contrast, in control, it is often necessary to incorporate human inputs in real time during policy execution, rather than through an offline process.
In this context, a key challenge is the design of interaction regulation mechanisms to incorporate online information arising from humans in a policy that is being executed by an autonomous agent. 
We introduce a control-theoretical framework to tackle this challenge and turn the framework into a control architecture. The framework relies on the minimization of the variational free energy, a central theme in, e.g., theoretical neuroscience~\cite{KF:09}, establishing a link between the free energy principle (FEP) and a certainty-equivalent adaptive optimal control scheme in the probability space. Our openly available architecture enables online integration of a policy generated by human preferences with the agent policy. The approach is experimentally validated on a testbed involving rovers autonomously navigating using onboard sensory information. While the robot executes its policy, a remote human, which receives the same sensory input via a virtual reality (VR) headset, conveys preferences to the robot through gestures. These preferences may induce cooperative or competitive interactions with the rover’s goal. Experiments demonstrate that preferences are effectively integrated into the rover’s policy, confirming the effectiveness of the interaction regulation mechanism.

The minimization of the variational free energy offers a unifying account across learning~\cite{eysenbach2022maximum,TB-AP-TM:24}, control~\cite{ET:09,Srivastava2023,TV-AO-HW:21,EG_HJ_CDV_GR:25}, Schr\"odinger bridges~\cite{Chen2015,doi:10.1137/20M1320195}, and neuroscience~\cite{TP-GP-KF:22}. Popular learning and control frameworks involve finding optimal policies by minimizing a free energy functional (see Sec.~\ref{sec:background} for the definition). The resulting optimal solution benefits, among other desirable properties, from implicit and explicit robustness guarantees~\cite{AS-HJ-KF-GR:26}. Policy computation via free energy minimization is increasingly seen as an appealing framework to design modular control architectures, often leveraging motor control tasks as benchmarks~\cite{prescott2023understanding,FR-VC-FB-GR:25}. Beyond modular architectures, this free energy framework is broadly recognized as a promising approach to design agents capable of performing complex tasks in largely uncertain environments. We refer to~\cite{lanillos2021activeinferenceroboticsartificial, GO-PL-GC:22} for a broader perspective, and survey, on the use of free energy minimization as a computational model to control autonomous agents. Our work is also related to the rich body of literature on cyber-physical systems with humans in the loop~\cite{MI-VG:19,SM-SH:17,SS-HL-AM-GR-RS:22} to adaptive optimal control, see, e.g.,~\cite{kohler2026certaintyequivalentadaptivempcuncertain} for a survey, and to information design for control~\cite{11615953,soleymani2024foundationsvalueinformationsemantic}.

\noindent{\bf Contributions.} We propose and validate a control-theoretic framework for interaction regulation. The framework integrates online human preferences into an agent's policy, allowing for both cooperative and competitive interactions. Grounded in variational free-energy minimization, the integration problem is formulated as optimal control in probability space, linking the FEP to an infinite-dimensional certainty-equivalent adaptive optimal control scheme. The resulting architecture integrates human and agent generated policies online and accommodates nonlinear, non-stationary, and stochastic dynamics and policies. We validate the approach experimentally through a virtual-reality interface in which a human intermittently communicates gesture-based preferences to a remotely located rover performing a lane-following task using onboard sensing. The results show that the proposed approach effectively incorporates human inputs into the rover's policy. Code, hardware details, experimental settings and a video are openly available at {\url{https://tinyurl.com/ae52ereu}}.

\section{Background}\label{sec:background}
We denote sets and operators by {\em calligraphic} characters, vectors by {\bf bold}, and the identity matrix of dimension $N$ by $\mathrm{I}_N$.
We let $\mathbb{K}$ be either $\R$ or $\Z$.
A random variable is denoted by $\bv{V}$ and its realization is $\bv{v}$. We denote the probability mass function (pmf, if the variable is discrete) or probability density function (pdf, for continuous random variables) by $p(\mathbf{v})$. We also let $\sD$ be the convex set of pdfs (pmfs). The expectation of a function $\mathbf{h}(\cdot)$ with respect to (w.r.t) the continuous random variable $\mathbf{V}$ is $\mathbb{E}_{p}[\mathbf{h}(\mathbf{v})] := \int_{\operatorname{supp} p} \mathbf{h}(\mathbf{v})\, p(\mathbf{v})\, d\mathbf{v}$, with $\operatorname{supp} p$ being the (compact) support of $p(\mathbf{v})$. If $\mathbf{V}$ is discrete, the integral is replaced by a sum. Whenever it is clear from the context we do not explicitly highlight the support in the integrals/sums. The joint pmf of $\mathbf{v}_1$ and $\mathbf{v}_2$ is $p(\mathbf{v}_1,\mathbf{v}_2)$ and the conditional pdf/pmf of $\mathbf{v}_1$ w.r.t. $\mathbf{v}_2$ is $p\left( \mathbf{v}_1\mid \mathbf{v}_2 \right)$. Given two pdfs/pmfs, $p(\bv{v})$ and $q(\bv{v})$, we say that $p(\bv{v})$ is absolutely continuous w.r.t. $q(\bv{v})$ if $\support p \subseteq \support q$. We denote this by writing $p \ll q$. The Kullback--Leibler divergence between $p(\mathbf{v})$ and $q(\mathbf{v})$ is $\DKL{p(\bv{v})}{q(\bv{v})} := \int_{\operatorname{supp} p} p(\mathbf{v}) \log \left( \frac{p(\mathbf{v})}{q(\mathbf{v})} \right) d\mathbf{v}$, if $\mathbf{V}$ is continuous. The integral
is replaced by a summation when $V$ is discrete.
The divergence quantifies the discrepancy of the pair $p(\mathbf{v})$, $q(\mathbf{v})$; $\DKL{p(\bv{v})}{q(\bv{v})}$ is non-negative, finite only if $p\ll q$ and $\DKL{p(\bv{v})}{q(\bv{v})}= 0$ if and only if $p = q$ almost everywhere~\cite{SK_RL:51}. 

\subsection{Minimization of the Variational Free Energy}
Given two pdfs/pmfs, $p(\mathbf{v})$ and $q(\mathbf{v})$, and a loss function, $l(\cdot)$, the variational free energy is defined as, see, e.g.,~\cite{9363495},
\begin{equation}\label{eqn:variational_free_energy}
\sF(p(\mathbf{v}),q(\mathbf{v})) := D_{\mathrm{KL}}\left(p(\mathbf{v}) \mid \mid q(\mathbf{v})\right) + \E_p\left[l(\mathbf{V})\right].
\end{equation}
The first term in~\eqref{eqn:variational_free_energy} is also known as statistical complexity and $q(\mathbf{v})$ is often termed as generative/time-series model. Several inference, learning and control problems involve (see Remark~\ref{rem:connections}) minimizing $\sF(\cdot,\cdot)$ w.r.t. its first argument. The following result summarizes key properties of the resulting optimization problem. The properties can be found under different technical conditions in, e.g.,~\cite{PG_MR_RW:14,Cammardella2023,ET:09, EG_GR:22}.
\begin{lemma}\label{lem:free_energy_optimization}
Given $q(\mathbf{v})$, consider 
\begin{equation}\label{eqn:free_energy_min}
  \min\nolimits_{
    p(\mathbf{v})\in\mathcal{D}
  } 
\;\mathcal{F}(p(\mathbf{v}),q(\mathbf{v})),
\end{equation}
with $\mathcal{F}(\cdot,\cdot)$ defined in~\eqref{eqn:variational_free_energy}. Then: (i) the problem is convex; (ii) the optimal solution is $p^{\ast}(\mathbf{v}) = \frac{q(\mathbf{v})\exp\left(-l(\bv{v})\right)}{\int q(\mathbf{v})\exp\left(-l(\bv{v})\right)d\bv{v}}$; (iii) the optimal value is {$-\log\int q(\mathbf{v})\exp\left(-l(\bv{v})\right)d\bv{v}$}.
\end{lemma}
\begin{remark}
Lemma~\ref{lem:free_energy_optimization} implies that the optimal solution of the problem in~\eqref{eqn:free_energy_min} is a pdf with an exponential kernel twisting the generative model~\cite{PG_MR_RW:14}. For discrete variables, the integral in the solution becomes a sum.
\end{remark}
\begin{remark}
Minimizing the KL divergence in~\eqref{eqn:variational_free_energy} amounts to minimizing the discrepancy between $p(\mathbf{v})$ and $q(\mathbf{v})$. The KL divergence in the objective can be thought of as a regularizer that biases the optimal solution of~\eqref{eqn:free_energy_min} towards $q(\bv{v})$. From a control-theoretic perspective, if
$q(\mathbf{v})$ represents a reference distribution, this term plays the role of a tracking error cost, penalizing deviations of  $p(\mathbf{v})$ from the reference.
\end{remark}

\section{Regulating Interactions via Free Energy Minimization}

\subsection{The Control Framework}\label{sec:policy_integration}
Consider an agent interacting with an environment. The environment transitions from state $\bv{x}_{k-1}$ to state $\bv{x}_k$ when the control input $\bv{u}_k$ is applied. The state space is denoted by $\sX\subseteq\mathbb{K}^n$ and the action space is denoted by $\sU\subseteq\mathbb{K}^m$. The possibly nonlinear and stochastic dynamics for the environment is $\plant{k}{k-1}$; we use the shorthand notation $\shortplant{k}{k-1}$ to denote $\plant{k}{k-1}$. We also let $\shortjointxu{k}{k-1} := \jointxu{k}{k-1} = \plant{k}{k-1}\policy{k}{k-1}$, where $\policy{k}{k-1}$ is a randomized policy. We use $\shortpolicy{k}{k-1}$ to denote this policy.

Given a time-step $k$ and the current state $\bv{x}_{k-1}$, $\plant{k}{k-1}$ characterizes the behavior of the environment, while $\policy{k}{k-1}$ represents the distribution from which the control input is sampled. Together, these distributions induce the closed-loop joint pdf $\jointxu{k}{k-1}$.

\begin{figure*}[t!]
    \centering
  
    \begin{minipage}[c]{0.44\textwidth}
      \centering
    
      \includegraphics[width=\textwidth]{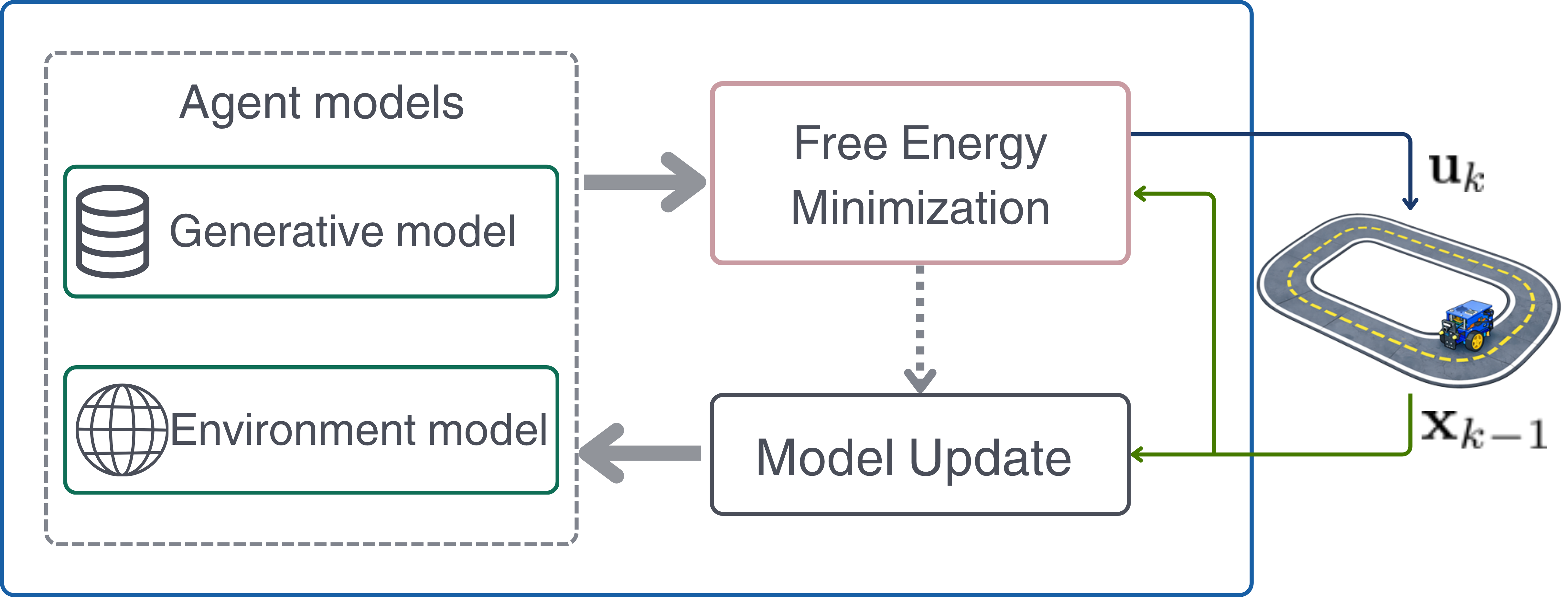}
    \end{minipage}
    \hfill
    \begin{minipage}[c]{0.54\textwidth}
      \centering
      \includegraphics[width=\textwidth]{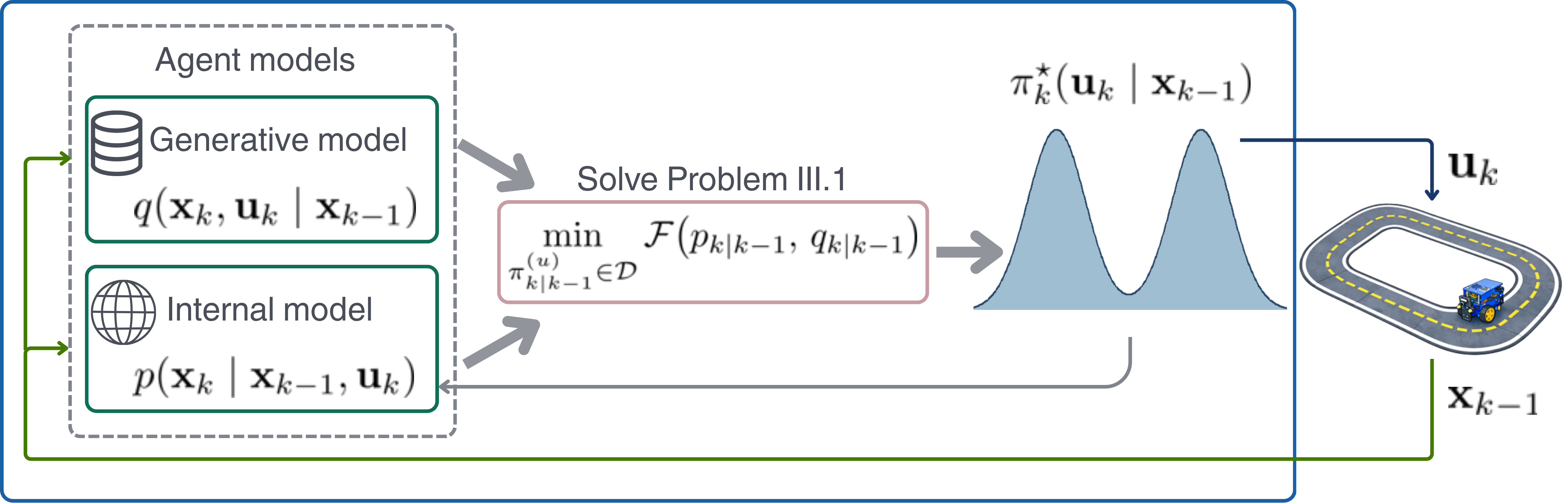}
    \end{minipage}
  
    \caption{An agent computing actions according to the FEP (left). Closed-loop data are used to update the environment model. Given this model, and a generative model, the free energy minimizing policy is computed. FEP as a certainty-equivalent adaptive optimal control scheme (right). The environment model and the generative model are internal models, with the former being learned from closed-loop data, either online or offline. The generative model can combine multiple pdfs/pmfs, capturing biases. In our framework these biases are generated from human preferences.     }
  
    \label{fig:FEP_control}
  \end{figure*}

We frame interaction regulation as a free energy minimization problem. Specifically, human preferences are integrated online into an agent policy by tackling an optimal control problem, where the free energy is minimized in the probability space. In the free energy, the loss encodes the agent (task-specific) cost, human preferences are encoded in the generative model. This yields the following:

\begin{problem}\label{prob:main}
At each $k$, given the current state $\bv{x}_{k-1}$ and:
\begin{enumerate}
\item an environment model $\plant{k}{k-1}$;
\item a preference-encoding generative model {$\shortrefjointxu{k}{k-1} := \refjointxu{k}{k-1} = \shortrefplant{k}{k-1}\shortrefpolicy{k}{k-1}$, with $\shortrefplant{k}{k-1} := \refplant{k}{k-1}$ and $\shortrefpolicy{k}{k-1} := \refpolicy{k}{k-1}$;}
\item $\mathcal{F}(\cdot,\cdot)$ defined as in~\eqref{eqn:variational_free_energy} with $l(\bv{x}_{k},\bv{u}_k) = \statecost{k} + \actioncost{k}$, where $\statecost{k}$, $ \actioncost{k}$ are the agent state/action costs.
\end{enumerate}
Find $\shortoptimalpolicy{k}{k-1}:= \optimalpolicy{k}{k-1}$ such that
\begin{equation}\label{eqn:main_problem}
  \shortoptimalpolicy{k}{k-1}
  \in
  \mathop{\mathrm{arg\,min}}\nolimits_{
    \shortpolicy{k}{k-1}\in\sD
  } \;
  \mathcal{F}\left(
    \shortjointxu{k}{k-1},
    \shortrefjointxu{k}{k-1}
  \right).
\end{equation}
\end{problem}
\vspace{-0.1cm}
An overview of the framework is shown in Fig.\ref{fig:FEP_control}.
\begin{remark}
In the cost of the problem in~\eqref{eqn:main_problem}, the loss function captures the task of the agent, which determines the agent policy if no human were present in the control loop. Instead, the KL divergence biases the optimal solution towards the generative model, and hence towards human preferences captured by the model.
\end{remark}
 
\begin{remark}\label{rem:connections}
The free energy minimization problem in~\eqref{eqn:main_problem} can be recast as a KL minimization problem, which naturally arises across learning and control, e.g., in the context of KL control, control as inference, and maximum diffusion/entropy reinforcement learning~\cite{TB-AP-TM:24,ET:09,HK-VC-MO:12}. This reformulation also links the problem to variational inference~\cite[Chapter 10]{KM:23} and fully probabilistic design. Originating from the seminal work in~\cite{MK:96}, the core ideas of this latter approach imply that the cost in the formulation can capture a cost-to-go that can be obtained via a backward recursion; constraints can be added to the optimal control problem~\cite{DG-GR:22}.
\end{remark}

The optimal solution of Problem~\ref{prob:main}, i.e., the free energy minimizing policy integrating human preferences (via the generative model) and the agent policy (via the agent cost) has an explicit expression that highlights the role of the agent cost and the generative model on optimal decisions. Standard derivations are omitted for brevity, as they follow established arguments. By the chain rule for the KL divergence, see, e.g.,~\cite{DG-GR:22}, $\mathcal{F} \left(\shortjointxu{k}{k-1},\shortrefjointxu{k}{k-1}\right)$ in~\eqref{eqn:main_problem} can be written as
\begin{equation}\label{eqn:reformulated_cost}
  \DKL{\shortpolicy{k}{k-1}}{\shortrefpolicy{k}{k-1}} + \E_{\shortpolicy{k}{k-1}}\left[ \tilde{l}_k(\bv{X}_{k-1},\bv{U}_k) \right],
\end{equation}
where $ \tilde{l}_k(\cdot,\cdot)$ is a combined loss given by
\begin{equation}\label{eqn:twisted_cost}
\DKL{\shortplant{k}{k-1}}{\shortrefplant{k}{k-1}}+ \E_{\shortplant{k}{k-1}}\left[\statecostexpectation{k}\right] + \actioncost{k}. 
\end{equation}
Hence, by means of Lemma~\ref{lem:free_energy_optimization} we have that, at each $k$, the integrated policy solving Problem~\ref{prob:main} is
\begin{equation}\label{eqn:optimal_integrated_policy}
\optimalpolicy{k}{k-1} = \frac{1}{Z_k} \refpolicy{k}{k-1}\exp\left(- \tilde{l}_k(\bv{x}_{k-1},\bv{u}_k)\right),
\end{equation}
where the normalizer $Z_k$ is $\E_{\shortrefpolicy{k}{k-1}}\left[\exp\left(- \tilde{l}_k(\bv{x}_{k-1},\bv{u}_k)\right)\right]$.

The expression in~\eqref{eqn:twisted_cost} is a free energy. In the context of the FEP, this functional is often minimized over past data w.r.t. $\plant{k}{k-1}$ to learn the environment model, which is subsequently used to compute the optimal policy via~\eqref{eqn:main_problem}; an exploration-promoting term can be included in the loss to facilitate learning. This results in a certainty-equivalent adaptive optimal control scheme in the probability space (Fig.~\ref{fig:architecture}, left). In applications (see, e.g., works surveyed in the introduction and references therein) the infinite-dimensional optimization is often tackled by resorting to policy/model parametrizations and/or Gaussian approximations. Remarkably, while the environment model is learned, the generative model is typically assumed to be known. In our framework, as detailed in the next sections, the generative model is obtained from human preferences.

\subsection{Control Architecture}\label{sec:architecture}
\begin{figure*}[t!]
   \centering
   \begin{minipage}[c]{0.7\textwidth}
      \centering
      \includegraphics[width=\textwidth]{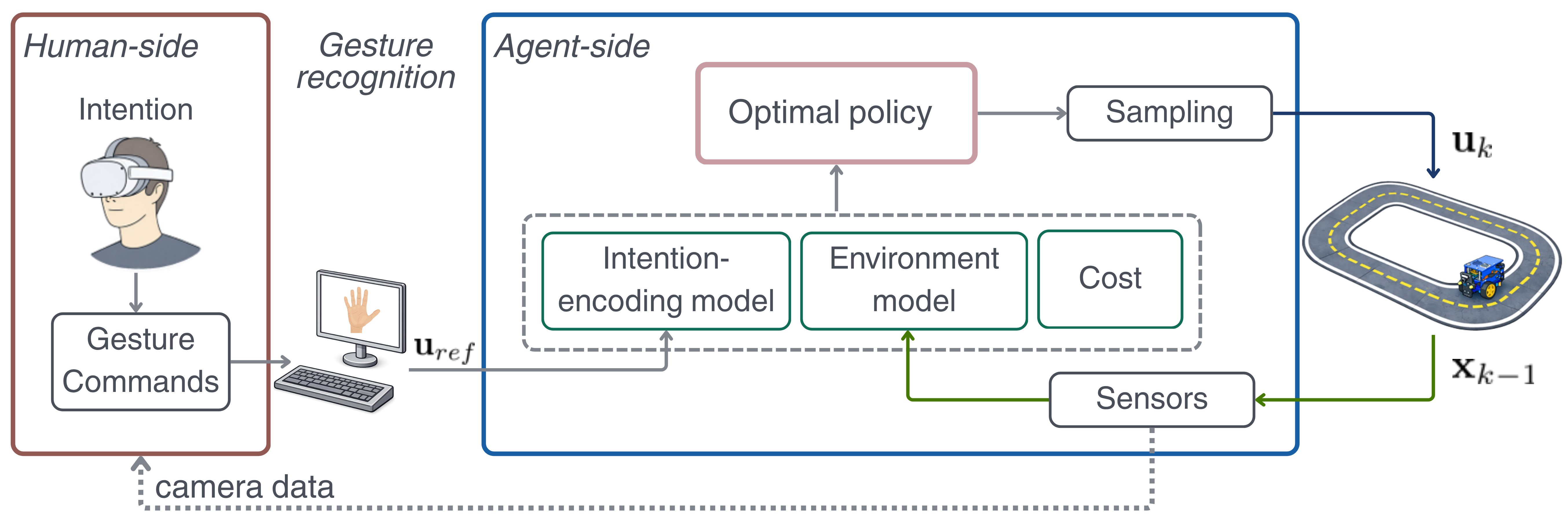}
     
    \end{minipage}
    \hfill
    \begin{minipage}[c]{0.28\textwidth}
      \centering
      \includegraphics[width=\textwidth, trim=0 6cm 0 11cm, clip]{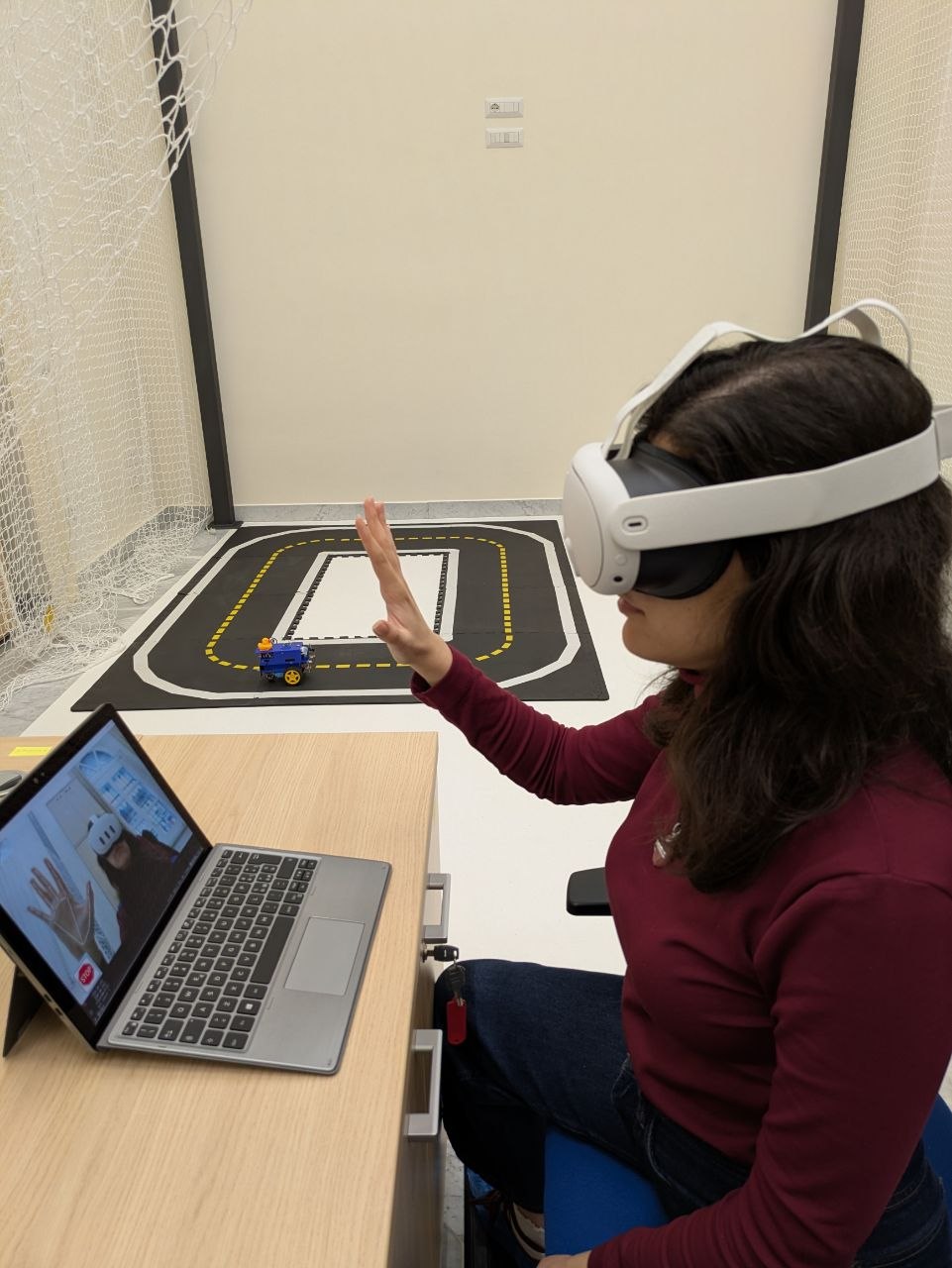}
      
    \end{minipage}
   \caption{Proposed architecture. Overview (left): the robot’s onboard camera stream is sent to a remotely located person. The person can generate high-level gestures. Gestures, once interpreted via a gesture recognition software, are mapped, on the agent-side, onto a generative model. On the agent-side, this model together with a model of the environment available to the agent and a cost that captures its goal, is used to build the integrated policy in~\eqref{eqn:optimal_integrated_policy}. A sampling routine is used to sample an action from the policy. Experimental setup (right): the robot is a Duckiebot and navigates on a circular map. The person is equipped with a MetaQuest VR headset and gestures are parsed via OpenCV running on a standard laptop.}
   \label{fig:architecture}
  \end{figure*}

The key functional components of the architecture, distributed across the agent (a rover equipped with a camera) and human sides, and their interactions with the environment are schematically illustrated in Fig.~\ref{fig:architecture} (left). The policy is computed on the agent-side. This free energy minimizing policy is computed according to~\eqref{eqn:main_problem} using the cost and environment model available to the agent, together with a generative model. The agent determines the action by sampling from the policy. The generative model used to compute the policy is obtained from the preferences provided by a remotely located person. The person receives the camera stream from the rover via a VR headset and can provide his/her preferences to the robot via gestures. These gestures are processed on the human-side of the architecture and the extracted preference $\bv{u}_{\text{ref}}$ is sent to the agent, where a preference/{intention}-encoding routine transforms the gestures into a generative model, $\refjointxu{k}{k-1}$. Next, we briefly discuss the blocks in the architecture (see \url{https://tinyurl.com/3v4yrf6d} for code, details, settings).

\noindent {\bf Sensors.}
The robot is equipped with onboard sensors that provide measurements of the environment.
In particular, a front-facing camera is used to estimate the lateral deviation $d_k$ and heading error $\phi_k$ w.r.t. the lane center.

\noindent{\bf Human-side.}
The remotely located human receives the robot’s onboard camera feed in real time through a VR headset. Hand gestures are captured and processed by an external computer using a feature-based gesture recognition algorithm. The recognized gesture is mapped to a preferential direction, which is transmitted to the robot.

\noindent{\bf Intention encoding routine.}
{Given the current lateral deviation and heading error, the preferred direction is embedded into a joint probability of states and actions, i.e., the intention-encoding generative model, $\refjointxu{k}{k-1}$.}

\noindent{\bf Environment model.}
{The model available to the agent is the kinematic model of the rover.}

\noindent{\bf Cost.}
{The cost captures the agent goal. In the experiments, we use a state cost that penalizes the increase in the lateral deviation and heading error.}

\noindent{\bf Optimal policy.}
The control policy is computed by implementing the expression in Eq.~\eqref{eqn:main_problem}, using the cost, environment and generative models described above. 

\noindent {\bf Sampling.}
At each time-step, the control input is sampled from the optimal policy and applied to the robot. In the current implementation, we pick the action with the highest probability. Alternative sampling strategies can be considered; an example is discussed in the following remark.

\begin{remark}
\label{rem: langevin}
At each $k$, given state $\bv{x}_{k-1}$, one can sample from~\eqref{eqn:optimal_integrated_policy} by simulating the Langevin dynamics~\cite{melanson2025thermodynamic} 
\begin{equation*}
d\mathbf{u}_k = -\nabla_{\mathbf{u}_k} \Phi_k(\bv{x}_{k-1},\mathbf{u}_k)\,dt + \sqrt{2}\,d\mathbf{w}(t),
\end{equation*}
where $\mathbf{w}(t)$ is a standard Wiener process and $\Phi_k(\cdot,\cdot)$ is
\[
\Phi_k(\bv{x}_{k-1},\mathbf{u}_k) = {\tilde{l}_k}(\bv{x}_{k-1},\bv{u}_k) - \log \refpolicy{k}{k-1}.
\]
While interesting {\em per se}, this approach may also enable sampling via thermodynamic hardware~\cite{melanson2025thermodynamic}.
\end{remark}
  
\section{Architecture Implementation}\label{sec:experiments}
\subsection{Set-Up}
The rover (Fig.~\ref{fig:architecture}, right) is a Duckiebot (\url{https://tinyurl.com/4n4p8x4v}) and navigates on a smooth rectangular road layout of $1.83$~m $\times$ $2.42$~m.
Each straight segment consists of a two-lane road (lane width $0.21$~m). The Duckiebot is a differential-drive robot equipped with a front-facing camera, an Inertial Measurement Unit, a time-of-flight sensor, and encoders. Computations are performed onboard using an embedded computing unit (NVIDIA Jetson Nano), enabling autonomous operation without external processing. The experimental setup also includes an external computer and a Meta Quest~3 headset (\url{https://tinyurl.com/5h4u2dy8}). The computer processes human gestures to generate a reference signal. As described below, this signal is processed on the robot to obtain a generative model. The headset allows the person to receive the robot's camera stream in real time.

\subsection{Control Pipeline}
We refer to our repository for detailed settings and parameter values. Let $v_k$ be the rover linear velocity and $\omega_k$ the angular velocity. In the current implementation, the control input $\bv{u}_k =[v_k,\omega_k]$ is discretized. The action space $\sU$ is $[0, v_\mathrm{max}] \times [-\omega_\mathrm{max}, \omega_\mathrm{max}]$, where $v_\mathrm{max}$, $\omega_\mathrm{max}$ are the maximum linear and angular admissible velocity values, respectively. Discretization is over a uniform grid of size $M_v \times M_\omega$, resulting in $M$ feasible control inputs.

Human gestures are mapped to a finite set of discrete preferred actions (\emph{Left}, \emph{Right}, \emph{Forward}, \emph{Stop}) and transmitted to the robot.
On the robot, these preferences are mapped to a desired input, $\mathbf{u}_{\mathrm{ref}}$, compatible with the robot kinematics. Specifically,
\emph{Left}, \emph{Right}, \emph{Forward} and \emph{Stop} are mapped onto $[0, \omega_\mathrm{ref}]$,
 $[0, -\omega_\mathrm{ref}]$,
 $[v_\mathrm{ref}, 0]$
 and $[0, 0]$, respectively.
 Unlike the other gestures, when the \emph{Stop} command is detected, the integrated policy is bypassed and the rover’s velocity is set to zero.
The vector $\mathbf{u}_\mathrm{ref}$ is updated asynchronously and held constant between sampling instants. Given this vector, $\refpolicy{k}{k-1}$ is computed as a truncated Gaussian centered in $\mathbf{u}_\mathrm{ref}$. That is, $\refpolicy{k}{k-1}
=
\mathcal{N}(\mathbf{u}_{\mathrm{ref}}, \Sigma^2_u)$,
where $\Sigma_u = \mathrm{diag}(\sigma_v, \sigma_\omega)$.
Also, $\refplant{k}{k-1}$ is a Gaussian over the state $\mathbf{x}_k = [d_k, \phi_k]$, where $d_k$ is the lateral deviation from the lane center and $\phi_k$ is the heading error. Namely, $
\refplant{k}{k-1}= \mathcal{N}(\boldsymbol{\mu}_q, \sigma^2 \mathrm{I}_2)$, where $\boldsymbol{\mu}_q$ is obtained from an offline-learned regression model trained on synthetic data generated from a simplified unicycle model. The generative model on the robot is then obtained as $\refjointxu{k}{k-1} = \refplant{k}{k-1}\refpolicy{k}{k-1}$.

\begin{remark}\label{rem:trust}
In the current implementation, the covariance matrices in the generative model are tunable parameters. As noted in, e.g.,~\cite{DG-GR:22}, higher variances may be associated to lower trust in the model. We reserve the problem of learning these parameters for future research.
\end{remark}
The model $\plant{k}{k-1}$ is given by the Gaussian $\plant{k}{k-1} = \mathcal{N}(\boldsymbol{\mu}_p, \sigma^2 \mathrm{I}_2)$, where $\boldsymbol{\mu}_p$ is obtained from the standard kinematic unicycle model, and $\sigma^2 \mathrm{I}_2$ captures the presence of sensors/actuators noise.

The agent goal, i.e., zero lateral deviation and zero heading error, is encoded in the state cost $\statecost{k}$, which consists of two terms so that $\statecost{k} = c^{(x)}_{\mathrm{err}}(\mathbf{x}_{k}) + c^{(x)}_{\mathrm{barrier}}(\mathbf{x}_{k})$. The first term is a cost that penalizes the lateral deviation from the lane center and the heading misalignment, i.e.,
\begin{equation}
c^{(x)}_{\mathrm{err}}(\mathbf{x}_{k})
=
\gamma_d \bar d_{k}^2
+
\gamma_\phi \bar \phi_k^2
+
\gamma_{\mathrm{c}}\, \bar d_{k}\,\bar \phi_{k},
\end{equation}
where $\bar d_k = \frac{d_k}{d_{\mathrm{scale}}}$, and $\bar \phi_k = \frac{\phi_k}{\phi_{\mathrm{scale}}}$, with $d_{\mathrm{scale}}$ and $\phi_{\mathrm{scale}}$ being normalization parameters used to balance the contribution of the lateral deviation and heading error in the cost. The second cost term, a barrier to the lateral deviation, is 
\[
c^{(x)}_{\mathrm{barrier}}(\mathbf{x}_k)
=
\gamma_{\mathrm{barrier}}
\left(
\frac{1}{\beta}
\log\!\left(
1+\exp\,\bigl(\beta(|\bar d_k|-d_{\mathrm{th}})\bigr)
\right)
\right)^2,
\]
where $d_{\mathrm{th}}$ is a threshold in the normalized lateral deviation determining the activation of the barrier penalty. The action cost $\actioncost{k}$ is set to zero in the experiments.
Given these elements, the policy is computed according to~\eqref{eqn:main_problem}  using the fact that $
D_{\mathrm{KL}}\!\left(
p_{k\mid k-1}^{(x)}
\;\|\;
q_{k\mid k-1}^{(x)}
\right)
=
\frac{1}{2\sigma^2}
\left\|
\boldsymbol{\mu}_p
-
\boldsymbol{\mu}_q
\right\|^2$. In the experiments presented next, sampling is performed by picking the action with the highest probability.

\section{Experimental Results}\label{sec:results}
\noindent{\bf Straight lane navigation.} The rover navigates along a straight lane section, aiming to keep both lateral deviation and heading error at $0$. However, it receives time-varying human preferences that may conflict with this goal. The experiment shows that the free-energy-minimizing integrated policy does not blindly execute these preferences, but selectively attenuates those that would lead the rover out of the lane while preserving the lane-following objective.

\begin{figure}[t!]
   \centering
      \includegraphics[]{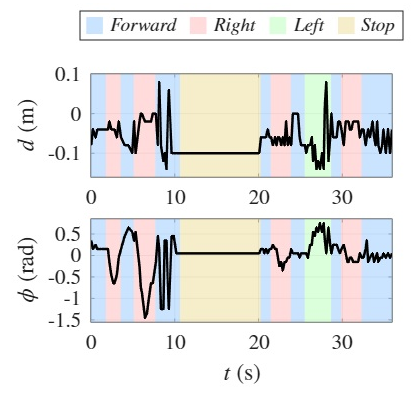}
   \caption{Background colors denote preferences received by the rover while it is navigating straight (colors online). Evolution of lateral deviation (top) and heading error (bottom). Despite conflicts between the agent task and human intentions, the robot maintains bounded deviation from the lane center.}
   \label{fig:exp2}
  \end{figure}

Fig.~\ref{fig:exp2} reports the time evolution of lateral deviation (top) and heading error (bottom). The experiments show that changes in the preferences induce transient variations in the heading errors. These are visible as peaks in $\phi $ and correspond to corrective steering actions performed by the robot to keep both $d$ and $\phi$ to $0$. Despite these transients, the lateral deviation remains bounded within a relatively small range throughout the experiment. In particular, throughout the experiment, $d$ has mean $-0.0538$~m and standard deviation $0.0407$~m, while $\phi$ has mean $0.0038$~rad and standard deviation $0.4160$~rad. This indicates that the robot maintains a limited average displacement from the lane center and does not accumulate systematic orientation bias over time. The larger variability observed in $\phi$, compared to $d$, suggests that the robot, due to the cost, primarily acts through heading corrections to compensate for conflicting commands. That is, orientation is actively adjusted to prevent lateral error accumulation. Even when the human input is inconsistent with the task, the robot does not drift outside the lane.

\noindent{\bf Full-lap.} We evaluate the integrated policy under consistent and conflicting human preferences during a full-lap lane-following task. The results show that the policy mediates between human input and the rover objective captured by its cost: task-consistent preferences are followed, while conflicting ones are down-weighted.

\begin{figure*}[t]
   \centering
   \begin{minipage}[c]{0.32\textwidth}
      \centering
      \includegraphics[width=\textwidth]{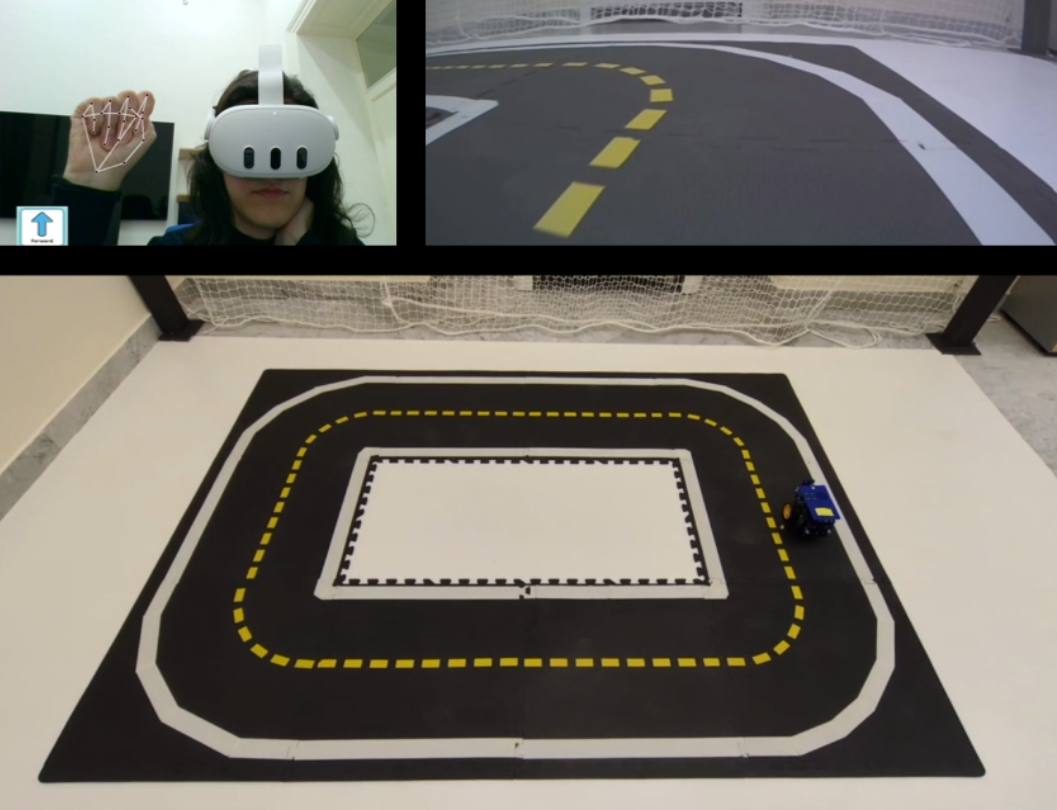}
     
    \end{minipage}
    \begin{minipage}[c]{0.32\textwidth}
      \centering
      \includegraphics[width=\textwidth]{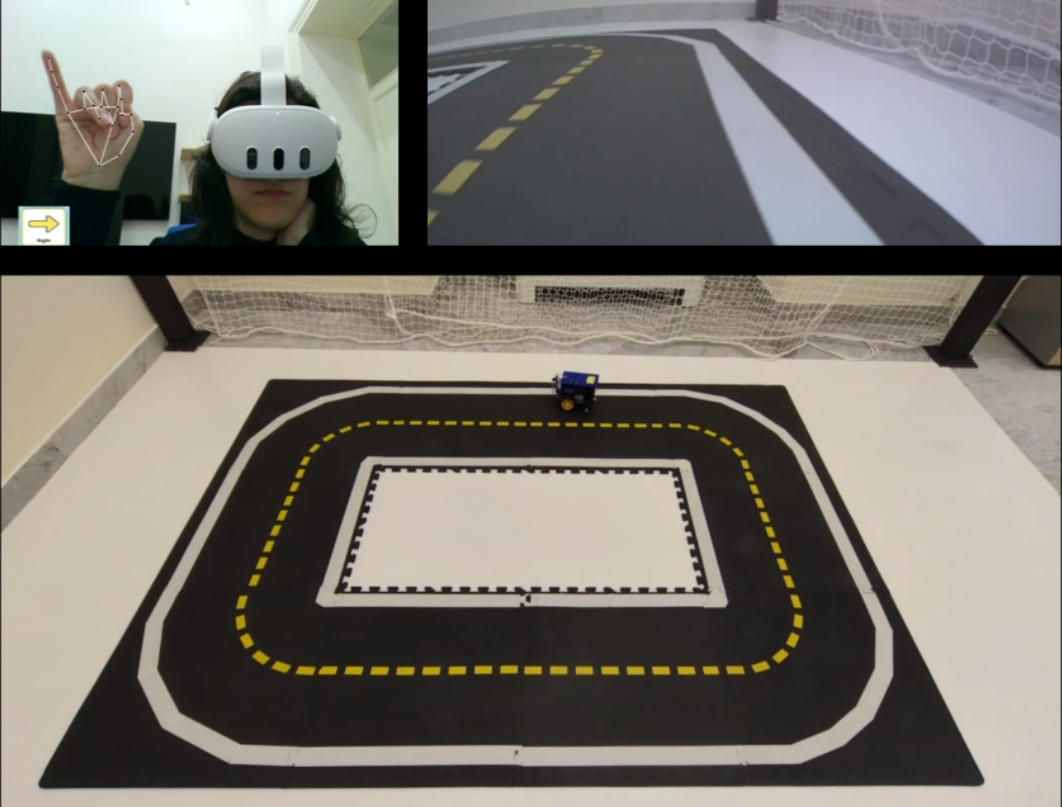}
      
    \end{minipage}
     \begin{minipage}[c]{0.32\textwidth}
      \centering
      \includegraphics[width=\textwidth]{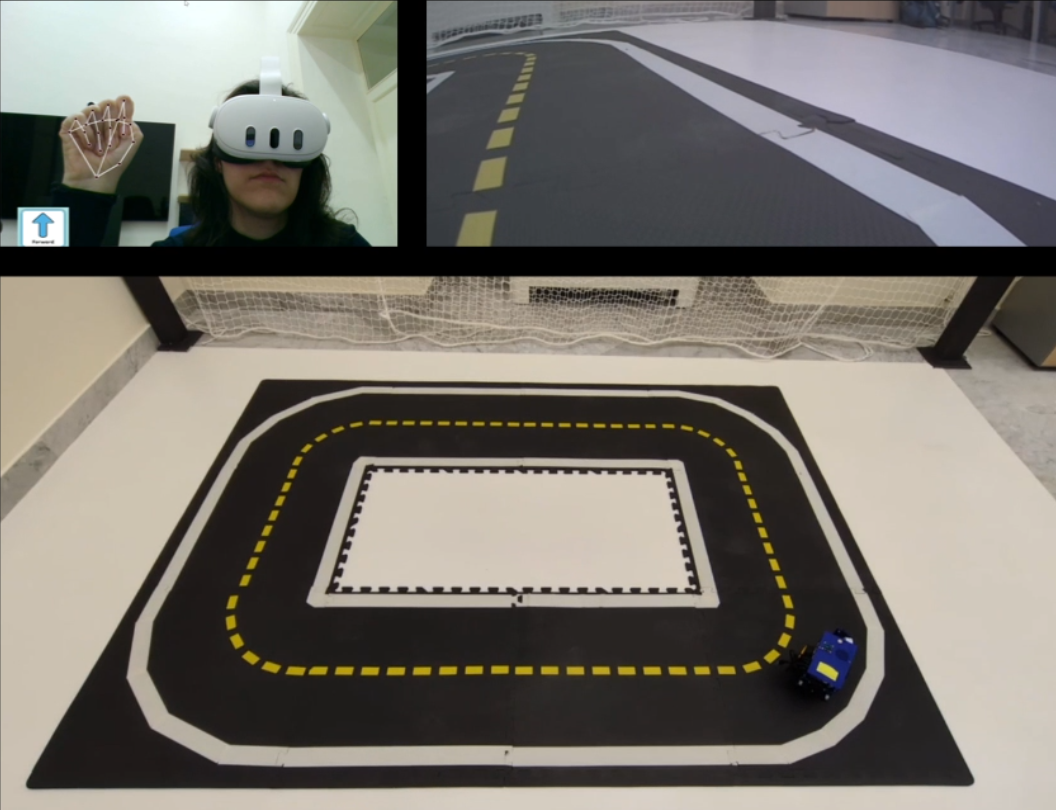}
    \end{minipage}
    \hfill
   \caption{Full-lap experiment. Three representative snapshots are shown. At $t=1$~s (left), the operator provides a task-consistent \emph{Forward} command on a straight segment, which the robot follows. At $t=10$~s (middle), the operator provides a \emph{Right} command inconsistent with lane following, and the robot maintains a straight trajectory. At $t=50$~s (right), the operator provides a \emph{Forward} command on a curved segment, and the robot turns to remain within the lane.}
   \label{fig:exp3}
  \end{figure*}

Fig.~\ref{fig:exp3} reports three representative snapshots illustrating different interaction scenarios between the human input and the rover from a video available at \url{https://tinyurl.com/ae52ereu}. The resulting behavior was obtained using a larger variance for $\refpolicy{k}{k-1}$, modeling a lower trust in human preferences (see Remark~\ref{rem:trust}). In this setting, at $t = 1$~s (left), the robot is on a straight segment and the operator provides a \emph{Forward} preference. In this case, the human preference is consistent with the task, and the resulting policy assigns high probability to actions aligned with the reference. As a consequence, the robot follows the command and maintains its position within the lane. At $t = 10$~s (middle), the robot is still on a straight segment, but the operator provides a \emph{Right} command. Although the command is admissible, it is not consistent with the lane-following objective. The policy assigns lower probability to such actions due to their higher predicted cost, and the robot continues to move straight, effectively rejecting the human input. Finally, at $t = 50$~s (right), the robot approaches a curved segment while the operator provides a \emph{Forward} command which is incompatible with the task, as maintaining a straight motion would lead the robot outside the lane. The policy compensates for this mismatch by selecting a turning action, allowing the robot to follow the curvature of the road.

\section{Conclusion}
We introduced a control-theoretical framework to regulate interactions between human and autonomous agents. The framework incorporates, online, human preferences into an autonomous agent policy. We turned the framework into an open control architecture and experimentally evaluated its effectiveness. In the experiments, a rover navigating using onboard sensory information interacted with a remotely located person. Sensory information was shared with the person via VR headsets and preferences, inducing both cooperative and competitive interactions, could be provided to the rover via gestures. Experiments demonstrated the effectiveness of our approach. 

The results presented here open a number of research questions. First, the architecture does not feature an online learning component and our future work aims to integrate policy computation under free energy minimization with online learning. This motivates analytical studies to assess closed-loop guarantees of the resulting dual control scheme. Second, we seek to develop the control architecture into a general-purpose framework capable of accommodating heterogeneous multi-agent settings, safety with control barrier functions, and diverse sensory information. Finally, exploiting the link with the Langevin equation (Remark \ref{rem: langevin}), we plan to deploy our method on thermodynamic hardware.

\noindent{\bf Acknowledgments.} The authors thank Prof. M. di Bernardo (Univ. of Naples Federico II and Scuola Superiore Meridionale) for his support and Dr. H. Jesawada (NYU Abu Dhabi) for initializing the set-up while at the Univ. of Salerno. Experiments were conducted at the CoRE Lab of the Scuola Superiore Meridionale.

\bibliographystyle{ieeetr}
\bibliography{cdc_refs}

\end{document}

%% file: packages.tex
\usepackage{graphicx}
\usepackage{lineno,hyperref}
\usepackage{tabularx} 
\usepackage{booktabs}
\usepackage{amsmath,amssymb,amsfonts}
\usepackage{stackengine}
\usepackage{graphicx}
\usepackage{amssymb}
\usepackage{eqnarray} 
\usepackage{amsmath}
\usepackage{mathalfa} 
\usepackage[dvips]{epsfig}   
\usepackage[ruled,vlined,linesnumbered]{algorithm2e}
\usepackage{psfrag}
\usepackage{amsmath}
\usepackage{stfloats}
\usepackage{amssymb}
\usepackage{dsfont}
\usepackage{xcolor}
\usepackage{bbm}
\usepackage{float}
\usepackage[prependcaption,colorinlistoftodos]{todonotes}
\usepackage{hyperref}
\usepackage{MnSymbol}
\usepackage{mathrsfs}
\usepackage{lscape}
\usepackage{longtable}
\usepackage{rotating}
\usepackage{multirow}
\usepackage{amsthm}
\usepackage{color}
\usepackage{xcolor}
\usepackage{url}
\usepackage{subfigure}
\usepackage{rotating}
\usepackage{hyperref}
\usepackage{tikz}
\usetikzlibrary{shapes.geometric, arrows}
\hypersetup{
    colorlinks=true, %set true if you want colored links
    linktoc=all,     %set to all if you want both sections and subsections linked
    linkcolor=blue,  %choose some color if you want links to stand out
}

\usepackage{fancyhdr}
\usepackage{chngcntr}

\usepackage{totcount}
\usepackage{atbegshi}

\usepackage[labelfont=bf]{caption} 
\usepackage{caption} % For customizing captions

\usepackage{pdflscape}
\usepackage{tabularx}

\usepackage{mwe}

\usepackage{hyperref}
\hypersetup{
    colorlinks=true,
    linkcolor=blue,
    filecolor=magenta,      
    urlcolor=cyan,
    pdftitle={Overleaf Example},
    pdfpagemode=FullScreen,
    }

%% file: my_commands.tex
\DeclareMathOperator{\support}{supp}

\newcommand{\sD}{\mathcal{D}}

\newcommand{\sF}{\mathcal{F}}

\newcommand{\sX}{\mathcal{X}}
\newcommand{\sU}{\mathcal{U}}

\newcommand{\R}{\mathbb{R}}
\newcommand{\E}{\mathbb{E}}
\newcommand{\Z}{\mathbb{Z}}

\newcommand{\bv}[1]{\mathbf{#1}}
\newcommand{\statecost}[1]{c_{#1}^{(x)}\left(\bv{x}_{#1}\right)}
\newcommand{\actioncost}[1]{c_{#1}^{(u)}\left(\bv{u}_{#1}\right)}

\newcommand{\statecostexpectation}[1]{c_{#1}^{(x)}\left(\bv{X}_{#1}\right)}

\newcommand{\Vx}[2]{V_{\alpha}\left(\bv{x}_{k-1},\bv{u}_k\right)}
\newcommand{\Wx}[2]{W_{\alpha}\left(\bv{x}_{k-1},\bv{u}_k\right)}
\newcommand{\Vxtilde}[2]{\tilde{V}_{\alpha}\left(\bv{x}_{k-1},\bv{u}_k\right)}
\newcommand{\Mx}[2]{M\left(\bv{x}_{k-1},\bv{u}_k\right)}

\newcommand{\jointxu}[2]{p_{{#1}}\left(\bv{x}_{{#1}},\bv{u}_{{#1}}\mid \bv{x}_{{#2}} \right)}
\newcommand{\shortjointxu}[2]{p_{{#1}\mid{#2}}}
\newcommand{\refjointxu}[2]{q_{{#1}}\left(\bv{x}_{{#1}},\bv{u}_{{#1}}\mid \bv{x}_{{#2}} \right)}
\newcommand{\shortrefjointxu}[2]{q_{{#1}\mid{#2}}}

\newcommand{\plant}[2]{p_{{#1}} \left(\bv{x}_{{#1}}\mid \bv{x}_{{#2}}, \bv{u}_{{#1}} \right)}

\newcommand{\shortplant}[2]{p_{{#1}\mid{#2}}^{(x)}}
\newcommand{\refplant}[2]{q_{{#1}} \left(\bv{x}_{{#1}}\mid \bv{x}_{{#2}}, \bv{u}_{{#1}} \right)}

\newcommand{\shortrefplant}[2]{q_{{#1}\mid{#2}}^{(x)}}

\newcommand{\policy}[2]{{\pi}_{{#1}}\left(\bv{u}_{{#1}}\mid \bv{x}_{{#2}} \right)}
\newcommand{\shortpolicy}[2]{{\pi}_{{#1}\mid{#2}}^{(u)}}
\newcommand{\refpolicy}[2]{q_{{#1}}\left(\bv{u}_{{#1}}\mid \bv{x}_{{#2}} \right)}

\newcommand{\shortrefpolicy}[2]{q_{{#1}\mid{#2}}^{(u)}}

\newcommand{\optimalpolicy}[2]{{\pi}^{\star}_{{#1}} \left(\bv{u}_{{#1}}\mid \bv{x}_{{#2}} \right)}
\newcommand{\shortoptimalpolicy}[2]{{\pi}_{{#1}\mid{#2}}^{(u),\star}}

\newcommand{\KL}{\text{KL}}

\newcommand{\DKL}[2]{{D}_{\KL}\left(#1\mid \mid #2 \right)}

\newtheorem{lemma}{Lemma}[section]

\newtheorem{problem}{Problem}[section]

\newtheorem{remark}{Remark}[section]

\usepackage{dsfont}

\fancypagestyle{SI}{
  \fancyhf{}

  \fancyfoot[C]{SI-\thepage}
}

\renewcommand{\thetable}{S-\arabic{table}}